# Nanoscale Heat Transfer at Contact Between a Hot Tip and a Substrate

Stéphane Lefèvre

Laboratoire d'Etude Thermiques, UMR CNRS 6608

Ecole Nationale Supérieure de Mécanique et d'Aérotechnique, 86960 Futuroscope Cedex

Sebastian Volz and Pierre-Olivier Chapuis

Laboratoire d'Energétique Moléculaire et Macroscopique, Combustion, UPR CNRS 288

Ecole Centrale Paris, 92295 Châtenay-Malabry

**Corresponding Author** :

Sebastian Volz, Ph.D.

EM2C-ECP, 92295 Châtenay Malabry, France

T : 33-1-4113-1049 F : 33-1-4702-8035,  volz@em2c.ecp.fr

**Abstract**

Hot tips are used either for characterizing nanostructures by using Scanning Thermal Microscopes or for local heating to assist data writing. The tip-sample thermal interaction involves conduction at solid-solid contact as well as conduction through the ambient gas and through the water meniscus. We analyze those three heat transfer modes with experimental data and modeling. We conclude that the three modes contribute in a similar manner to the thermal contact conductance but they have distinct contact radii ranging from 30nm to 1micron. We also show that any scanning thermal microscope has a 1 to 3 microns resolution when used in ambient air.

**Nomenclature**:

A: accommodation coefficient

a: thermal diffusivity ($m^2.s^{-1}$)

b: contact radius (m)

$C_{v,p}$: heat capacities ($J.kg^{-1}.K^{-1}$)

E: Young's modulus: (Pa)

e: film thickness (m)

F: force between the tip and the surface (N)

G: thermal conductance ($W.K^{-1}$)

H: hardness (Pa)

h: heat transfer coefficient ($W.m^{-2}.K^{-1}$)

I: electrical current (A)

L: half length of the rhodium-platinum wire (m)

p: probe perimeter (m)

Pr: Prandtl number

R: electrical resistance ($\Omega$) – Radius (m)

S: rhodium-platinum wire surface ($m^2$)

T: temperature (K)

V: voltage (V)

v: mean velocity of molecules in air ($m.s^{-1}$)

x: coordinate along the Pt-Rh wire axis (m)

$x_0$ coordinate on the Pt-Rh wire surface (m)

$y_0$: coordinate on the Pt-Rh wire surface (m)

z: tip altitude (m)

$z_0$: coordinate on the Pt-Rh wire surface (m)

**Greek symbols**:

$\alpha$: temperature coefficient ($K^{-1}$)

$\theta$: temperature amplitude (K)

$\gamma$: heat capacities ratio

$\lambda$: thermal conductivity ($W.m^{-1}.K^{-1}$)

$\rho$: electrical resistivity ($\Omega.m$)

**Subscripts**:

A: air

C: total contact conductance - probe curvature radius

Eq: contact and sample conductances in series

P: probe

S: solid-solid contact

W: water meniscus

x: ellipse small axis

y: ellipse large axis

**Keywords**:

Scanning Thermal Microscope, Nanoscale Heat Transfer.

Thermoelectric energy conversion was improved by a factor of 2 in the year 2001 by using nanostructured materials [1], the future of data storage is believed to rely on nanoscale heating [2], and nanomaterials are to be used for building insulation. Those examples emphasize the key role of heat transfer in nanotechnologies, especially regarding to the energy field. A review on the scientific challenges in microscale heat transfer can be found in several references [3,4].

A complex heat transfer issue is clearly encountered when predicting the heat flux between a hot tip and a sample. In the ambient air, heat conductions through solid-solid contact, through the gas and through the water meniscus are combined as illustrated by Figure 1. The tip sample contact conductance $G_c$ is defined as the sum of the three thermal conductances:

$$G_C = G_S + G_A + G_W. \qquad (1)$$

The thermal transport is governed on the quantitative and geometrical point of views by those three contributions. Those contributions can not be ignored when using the scanning probe microscopes. The spatial extension of the thermal interaction between the tip probe and a nanostructure is crucial. The flux value is also a keypoint when a tip heating is used to lower the local coercitive magnetic field or to melt a substrate in the case of data storage.

Previous works have reported a detailed analysis of the thermal mechanisms at point contact between a thermocouple tip and a hot substrate [5]. The air contribution is found to be dominant because the tip cantilever is heated through air. We propose to use a hot tip so that the measurements are not dependent on the temperature distribution on the sample surface.

Gomes et al [6] suggested that the water meniscus might be the dominant heat transfer mode but that this contribution should depend on the sample thermal conductivity.

In our previous papers [7-9], we identified the contact radius as being 1 micron when the tip temperature is larger than 100°C and about 200nm when it is lower. We presume that the change in the contact radius produces a change in the modes contributions.

We use a Scanning Thermal Probe Microscope to provide quantitative data for the thermal contact conductance and the contact radii of the three main modes. A presentation of the microscope is provided in the first paragraph. We address the solid-solid thermal interaction in the second part. The water meniscus contribution is studied based on a simple modeling in the third part. In the last section, the air contribution is analyzed with experimental and modeling tools.

**1. The SThM based on a Hot Tip**

The basis of most SThMs is the Atomic Force Microscope. Its principle is to maintain a constant force between a tip and a sample. A piezoelectric crystal controls the force by monitoring the height of the tip cantilever. The piezocrystal voltage is then directly related to the sample topography when the probe scans the surface. The original function of those systems was to provide the samples topography with the atomic resolution. Those devices were rapidly developed to also measure a large variety of local properties –magnetic, electric, elastic, …-. And in 1986, K. Wickramasinghe [10] proposed to mount a thermocouple tip in a conventional AFM. While the temperature was the feedback signal to control the tip height, it is until now used to measure the local temperature [11] when the tip is brought in contact with the surface. Those techniques however require an external heating [12] and the knowledge of the sample geometry to provide local thermal properties.

Our probe consists in an electrical resistance that is thermally controlled through Joule dissipation. The probe temperature is directly deduced from the probe electrical resistance. Those 'active' tips also measure the local thermal conductivity without any external heat source: the input current is controlled so that the tip temperature is maintained to a constant value, the feedback current then reflects the capacity of the sample to conduct heat. In the present paper, we used the 3-omega technique to measure the tip temperature [13,9]. An AC current at frequency ω is heating the tip at the frequency 2ω. The tip electrical resistance is linearly dependent to the temperature amplitude $\theta_{2\omega}$, $R=R_0(1+\alpha\theta_{2\omega})$ where α is the temperature coefficient. The tip voltage V=R(2ω).I(ω) therefore includes a thermal component at 3ω :

$$V_{3\omega} = \frac{I_0}{2} R_0 \alpha \theta_{2\omega}, \qquad (2)$$

where $I_0$ is the current amplitude. This technique allows us to remove the dependence of the measurement to the ambient temperature and ensures a high signal to noise ratio.

As illustrated in Figure 2, the tip is made of a wollaston wire of diameter 75microns and shaped as a tip. The silver coating is removed at the tip-sample contact to uncover the platinum/rhodium wire of diameter 5 microns. Due to the Joule heating, the temperature profile in the tip is parabolic. The temperatures at both ends are set to the ambient because the silver is assimilated to a heat sink. The detailed solving of the thermal problem is proposed in references [7,9]. The expression of the probe-sample conductance $G_{eq}$ including the contact $G_C$ and the sample $G_S$ contributions writes:

$$\frac{1}{G_{eq}} = \frac{1}{G_C} + \frac{1}{2\pi\lambda_S b}, \qquad (3)$$

where $\lambda_S$ and b are the sample thermal conductivity and the thermal contact radius. $G_{eq}$ is related to the measured temperature through:

$$\overline{\theta_{2\omega}} = \frac{1}{L}\int_0^L \theta_{2\omega}(x)dx \\ = \frac{J_0}{Lm^3} \frac{A.G_{eq} + B.G_P mL}{(\exp(2mL)-1)G_{eq} + (1+\exp(2mL))G_P mL}, \qquad (4)$$

where $J_0 = \dfrac{\rho I_0^2}{2\lambda_P S_P^2}$, $\rho$ being the probe electrical resistivity, $\lambda_P$ and $S_P$ the probe thermal conductivity and section. $G_p$ represents the probe conductance. The A and B coefficients are defined as:

$$A = \left(-2 - mL + 4\exp(mL) - 2\exp(2mL) + mL\exp(2mL)\right), \tag{5}$$

$$B = \left(1 + mL - \exp(2mL) + mL\exp(2mL)\right), \tag{6}$$

where L is the half length of the platinum wire. $m^2 = \dfrac{hp_P}{\lambda_P S_P} + \dfrac{2i\omega}{a_P}$ represents the probe fin parameter where h is the heat transfer coefficient between the tip and the ambient. $p_p$ and $a_p$ are the probe perimeter and thermal diffusivity.

## 2. The solid-solid and water meniscus contact conductances

The contact between two bodies is achieved through constrictions and spacing including gas and water. A thermal resistance appears due to the lower thermal conductivity of air and water but also due to the change of the flux lines that preferably pass through the constrictions. As illustrated in figure 1, the solid-solid contact between the tip and the sample is described by the same morphology. Consequently, we use the same model to describe the dependence of the conductance to the applied force [14]:

$$G_S = C.F^n = C'.\Delta I^n, \tag{7}$$

where C and C' are coefficients, F represents the force applied by the tip on the sample and $\Delta I$ is the current that controls the piezoelectric crystal extension. This current is proportional to the force. The literature [14] proposes a value of n between 0.63 and 0.99. Increasing the force smashes the constrictions and increases their conductance as well as the overall solid-solid contact conductance. We shall assume that the tip shape is not modified on the microscale so that the force dependence of the total conductance writes:

$$G_{eq} = \frac{2\pi\lambda_S b\left(C'\Delta I^n + G_A + G_W\right)}{2\pi\lambda_S b + C'\Delta I^n + G_A + G_W}. \tag{8}$$

Thermal mapping were performed on the surface of an Hafnium sample under different forces. The total conductance was identified based on Eqs. (4-6) and averaged on the surface. Figure 3 reports the comparison between experimental results and the prediction of Eq. (7). The fit provides $G_S = 6.8 \cdot 10^{-5}$ W.K$^{-1}$, $G_A + G_W = 9.8 \ 10^{-6}$ W.K$^{-1}$, $n=1$ and $C' = 2.1 \ 10^{-7}$ W.K$^{-1}$.A$^{-1}$. A change in the sample modifies the solid-solid contact conductance through its contact radius. Those are modelled through the Hertz law in the elastic domain:

$$b_s = \left(\frac{6RF}{E}\right)^{1/3}, \tag{9}$$

$R_P$ being the Pt-Rh wire radius, and in the plastic domain:

$$b_s = \left(\frac{4F}{\pi H}\right)^{1/2}, \tag{10}$$

where $R_C = 5\text{-}15\mu m$ is the tip curvature radius, E the Young's Modulus and H the hardness. An estimation of $b_{s\text{-}s}$ with typical values for E and H is 20nm. The power laws 1/3 and 1/2 emphasize a low sensitivity of the radius $b_{s\text{-}s}$ to the materials. We therefore believe that a variation of $G_s$ in the range of 0-5 $10^{-6}$ W.K$^{-1}$ is a reasonable general estimation. A 8nA current is usually applied when using the SThM tip so that $G_s \approx 1.7 \ 10^{-6}$ W.K$^{-1}$. This is 17% of the total conductance as learnt from the value of $G_A + G_W$. The reference value of $\lambda_S = 23$ W.m$^{-1}$.K$^{-1}$ also leads to a mean contact radius b = 740 nm >>20nm. We deduce that the air and the meniscus conductances might have contact radii much larger than the solid-solid one.

In ambient air, the hygrometric rate ranges from 35% to 65% and water molecules are adsorbed on samples surfaces. In AFM measurements, this water film is observed when measuring the cantilever deflection when the voltage of the piezoelectric crystal varies. When approaching the surface, the tip is brought down by capillarity forces. The film thickness can be estimated to 0.25-1nm from this signal. 0.25nm is the water molecule radius. The water meniscus was indicated as the main heat transfer channel in several studies [6]. We propose an estimation of the meniscus conductance including the tip geometry. The tip is assimilated to a half-tore and the sample as a plane surface. The tore equation has to include the curvature radius of the Pt-Rh wire $R_C$ and the wire radius $R_P$:

$$z_0 = R_C + R_P - \sqrt{R_C^2 + R_P^2 - y_0^2 - x_0^2 + 2R_C\sqrt{R_P^2 - x_0^2}} \quad . \tag{11}$$

Equation (11) relates the altitude $z_0$ of the tip to the coordinates $x_0$ and $y_0$ of a point M on the sample surface. $z_0$ also represents the meniscus thickness under the tip when $z_0<e_w$, $e_w$ being the film thickness. The heat transfer is assumed to be vertical so that a heat transfer coefficient can be defined as:

$$h(x_0, y_0) = \frac{\lambda_w}{z_0(x_0, y_0)}. \tag{12}$$

The thermal conductivity of water $\lambda_w$ is set to 0.61 W.m$^{-1}$.K$^{-1}$. The water conductance then writes:

$$G_W = \int_\Sigma h(x, y) dx.dy \tag{13}$$

where $\Sigma$ is the surface defined by $(x_0, y_0)$ points for which $0.25nm<z_0<e_w$. The water conductance ranges from $10^{-6}$ W.K$^{-1}$ for a one molecule thick film to $3.10^{-5}$ W.K$^{-1}$ when $e_w=1nm$ – 4 molecules thick film - as reported in figure 4. This contribution remains of the same order of magnitude than the solid-solid contact conductance. The contact radii as a function of $e_w$ are derived from analytical calculations and presented in Table 1. They are one order of magnitude larger than the solid-solid contact conductance.

### 3. Conduction through air

We will show that the thermal signal varies far before the tip is brought in contact with the sample. The radiation conductance can be overestimated to 10nanoW.K$^{-1}$ which is clearly negligible. We therefore presume that conduction through air is the key channel. The diffusive, slip and ballistic regimes of heat transfer were already modelled [15] to describe the rarefied gas effect on energy exchange between the tip and the sample. A 1D vertical conduction is also assumed. A local heat transfer coefficient is modelled as:

$$h(x_0, y_0) = \frac{\lambda_A}{z_0}, \qquad (14)$$

in the diffusive regime when $z_0$ is much larger than air mean free path (MFP) $\Lambda$=100nm. In the slip regime when $z<100\Lambda$, molecules temperature is strongly different from the one of the sample surface when colliding it. The heat transfer coefficient writes:

$$h(x_0, y_0) = \frac{\lambda_A}{z_0 + 2\left[(2-A)\gamma / A(\gamma+1)\Pr\right]\Lambda}, \qquad (15)$$

where A=0.9 is the rate of the molecule energy left to the surface, $\gamma=C_p/C_v$=1.4 and Pr=0.7 is the Prandtl number in air. This complex expression fits the diffusive regime when $z_0 \gg \left[(2-A)\gamma / A(\gamma+1)\Pr\right]\Lambda$ and also to the ballistic regime when air molecules do not collide between themselves. In this case, MFP is set to $z_0$ and:

$$h(x_0, y_0) = \frac{C_v.v.z_0/3}{z_0\left[1 + 2(2-A)\gamma / A(\gamma+1)\Pr\right]}, \qquad (16)$$

where the kinetic expression of the thermal conductivity $\lambda_A = C_v.v.z_0/3$ was introduced with the mean molecule velocity v. In the ballistic case, $h_A$ is not $z_0$ dependent anymore. The air conductance is then derived from expression (16).

We apply this modelling to the specific shape of the wollaston probe. The results are compared to experimental signals obtained when the tip altitude ranges from 150 microns to contact.

The landing of the tip on the surface starts at altitude 150 microns. The maximum dilatation of the piezoelectric crystal is of a few microns. We therefore use the vertical displacement generated by the motorized screw. This screw performs the tip approach before contact in a conventional AFM imaging. To measure the vertical displacement, another z probe was put in contact with the screw head. $G_{eq}$ was derived from Eq. (4) and from the 3ω tip voltage.

A silver sample was used to keep the air resistance larger than the sample one so that $G_{eq}=G_A$. Figure 5 reports the thermal resistance $1/G_{eq}$ versus the altitude z.

Beyond 20 µm, we presume that a convective regime is observed in Figure 5(a), i.e. lifting Archimedes forces become larger than viscous forces. Heat conduction mostly occurs in the viscous layer at the probe vicinity. The thickness of the viscous layer is approximated from $\lambda_A$ and the heat transfer coefficient h between the Pt-Rh wire and the ambient [9] $\frac{\lambda_A}{h}=25\mu m$.

This thickness precisely corresponds to the limit of air conduction regime where $R_{eq}$ is linearly dependent to z.

A deviation to this linear dependence appears in Figure 5(b) below z = 1µm. This trend is relevant to the slip regime and the small plateau when z<300 nm might correspond to the ballistic regime. Just before contact, the air conductance $G_A$=2.5 µW.K$^{-1}$ and $G_A$=2µW.K$^{-1}$ when the deviation appears. Consequently, the slip and ballistic regimes might contribute to 20% of the conductance through air when the tip is in contact.

The intersection between a plane of altitude Λ=100nm and the tore representing the tip is an oval. Its mean radius b can be defined with the two axis lengths $b_x$=2.5µm and $b_y$=9µm according to:

$$b=\sqrt{\frac{b_x^2+b_y^2}{2}} \ . \tag{17}$$

We obtain a very large value for b= 1.3µm. According to the previous modelling, this radius defines the surface on which the tip heating through air is governed by the efficient ballistic regime. The true value of $b_A$ has to be larger than 1.3µm.

Between z=1µm and 25µm the diffusive regime is observed. The 3 regimes model (3RM) assumes a diffusive behaviour above z=10 µm only. Understanding that the diffusive behaviour might be relevant on a wide z range, we perform a 3D finite elements modelling (FEM) of the tip-sample interaction based on the Fourier heat conduction equation. We neglect the enhancement of heat flux in the ballistic area because the FEM predictions show that the heat transfer in the ballistic area is much less than the total heat transfer. The tip is assimilated to an ellipsoid with small and large axis $b_x$ and $b_y$. Joule heating generates a parabolic temperature distribution in the probe. Therefore, the area of the probe that is in contact with the sample is the hottest part and the one that contributes most to heat transfer. The sample and the air are simulated by two adjacent cubes of 100µm in edge. The temperature on the ellipsoid boundaries is set to 400K, the ellipsoid is positioned in air at various altitudes from the sample. The temperatures of the outer boundaries of the two cubes are set to 300K. $G_{eq}$ is the ratio between the heat flux crossing the whole sample/air interface and 100K. We checked that changing the ellipse temperature would not change the value of $G_{eq}$. Our simulations includes about 60 000 elements. The mesh is refined around the ellipsoid volume.

Figures 6(a) and 6(b) reports the comparison between experimental measurements (diamonds), the simplified 3RM (continuous grey line) and the finite element modelling (black triangles). Discrepancies between 3RM and measurements mostly occur between 2 and 15 microns. This confirms that the slip regime is introduced at too high altitudes in the 3RM. The slip regime underestimates the conductance as shown in Figure 7. But there is a good agreement when z<1µm. The assumption of vertical conductance used in the 3RM is valid in the range of small z values indeed. The FEM and the experimental data have the same evolution but the FEM overestimates the measurements values by a factor of 2 when z<0.2μm and by 0.5 μW.K$^{-1}$ for higher altitudes. We emphasize that the diffusive conductance is higher than the ballistic one when z<Λ as indicated by Figure 7. The Fourier heat exchange coefficient follows a 1/z law and is diverging when z goes to zero whereas the ballistic conductance is constant. Of course, using the Fourier law when z<Λ is not physically relevant. This however explains why the FEM predictions are drastically overestimating measurements when z<0.2μm.

The thermal conductivity of the sample is dependent on the roughness and surface oxydation, it therefore might be lower than the reference value of 428 W.m$^{-1}$.K$^{-1}$ for Ag. The FEM calculation for $\lambda_S$=0.1 W.m$^{-1}$.K$^{-1}$ was performed and reported in Figure 6a (empty triangles). The FEM values then match measurements better when z>20μm.

The FEM data do not follow the linear behaviour when z<3μm as seen in Figure 6(b). The vertical heat transfer coefficient approximation 1/h∝z is yet more reliable near contact. We presume that heat flux from the surface Σ of the ellipsoid becomes non-homogeneous when z is small. This behaviour is z-dependent. The Taylor expansion of h∝1/z-(z(dΣ)-z)/z$^2$ where dΣ is the surface element on Σ proves that h is much z(dΣ) dependent when z is of the order of z(dΣ)-z, i.e. about $R_P$=2.5μm. The deviation observed in the experimental data might also be due to this effect so that the slip regime is likely to start for even lower altitude than 1 μm.

The predominance of the heat diffusion in air on the contact conductance implies that the contact radius and the microscope resolution depends on the sample thermal conductivity. The flux lines spread when the sample thermal conductivity decreases. We computed the spatial distribution of the heat flux crossing the sample surface by using our FEM. The tip height is 20nm so that no solid-solid heat conduction is involved. Figure 8 reports a slight difference in the flux distributions when $\lambda_s$ ranges from 100 to 5 W.m$^{-1}$.K$^{-1}$. But the maximum flux values then decreases by a factor of 5 when $\lambda_s$ reaches 0.1 W.m$^{-1}$.K$^{-1}$. The contact radius can be identified as the radius for which the heat flux density reaches 50% or 90% of its maximum value. The insert of Figure 8 shows that the radius increases by a factor of 2 (90%) or 25% (50%) when the sample thermal conductivity decreases to the air thermal conductivity. In those conditions, the range of radius values is 1.5-3.3µm (90%) and 4 -5.4µm (50%). A value of 1µm for b was obtained in previous works [7] from experimental data when the tip temperature is higher than 100°C. In those conditions, the meniscus disappears and air conduction becomes predominant. We therefore believe that our estimation of b remains reasonable.

## 4. Conclusion

We have presented experimental and modelling results to understand and quantify the heat transfer mechanisms between a micrometer tip and a sample surface. The conventional law was retrieved for the heat transfer due to the solid-solid contact. Values of 1.7 µW.K$^{-1}$ and 20nm were obtained for the thermal conductance and the radius. Conduction in the meniscus was estimated from the probe geometry. A 1 to 30 µW.K$^{-1}$ was obtained for water film thicknesses as small as 4 water molecules. The order of magnitude of the radius is 100nm. Air conduction between the tip and the sample was studied in details. A thermal conductance of 2.5 µW.K$^{-1}$ and we proved that the corresponding radius ranges from 1.5 to 3 µm depending on the sample thermal conductivity. As shown in Table 2 and in Figure 9, the three heat transfer modes have similar contributions with a predominance of the water meniscus depending on the hygrometric rate. The radii have very different order of magnitudes. Working with a hot tip removes the meniscus and the tip contact radius then becomes of the order of the micron: a nanoscale contact requires working in vacuum.

**CAPTIONS**

**Table 1:** Contact radius $b_W$ corresponding to heat conduction in the water meniscus for different water film thickness.

**Table 2**: Thermal conductances and radii for the four heat transfer modes involved in the tip-sample heat transfer.

**Figure 1**: Schematic of the probe-sample interaction including conduction through air, through the water meniscus and through the solid-solid contact.

**Figure 2**: Scanning Electronic Microscope image of the thermal probe. The Wollaston wire is a silver coating 75 microns in diameter and a Pt-Rh core 5 microns in diameter. The mirror ensures the laser reflection to control the tip deflection.

**Figure 3**: Thermal conductances of the contact and the sample versus the force applied by the tip on the sample.

**Figure 4**: Thermal contact conductance through the water meniscus versus the meniscus thickness.

**Figures 5(a) and 5(b)**: Thermal resistance of the contact and the sample versus the tip altitude. Figure 5(a) reveals a convective regime when z>20μm and a linear regime corresponding to conduction in air when z<20μm.

**Figures 6(a) and 6(b)**: Comparison between the measured conductance and the predicted ones. The modeling is based on a 3D finite element method scheme (FEM) and a simplified 3-regimes description (model). The figure 6(b) reports the resistance versus altitude. The linear regime corresponds to conduction in air. The three approaches predict the same thermal conductance through air as shown by the extrapolation for z=0.

**Figure 7**: Heat transfer coefficients in the 3 regimes model.

**Figure 8**: Flux versus radius (small ellipse axis direction) when the tip is in contact and for different values of sample thermal conductivities. The insert reveals that the contact radius due to air conduction may vary with the sample thermal conductivity by a factor of 2.
.

Table 1:

| Film Thickness $\delta_W$ (nm) | $b_W$ (nm) |
|---|---|
| 0.25 | 100 |
| 0.5 | 140 |
| 1 | 200 |

Table 2:

| Heat Transfer Mode | Conductance ($\mu$W.K$^{-1}$) | Contact Radius b (nm) |
|---|---|---|
| Radiation | $\approx 10^{-3}$ | - |
| Solid-solid | 0 – 1.8 | $\approx 20$ |
| Conduction through air | $\approx 2.5$ | 1000 – 3000 |
| Water Meniscus | 5 - 30 | 100 - 200 |

Fig 1:

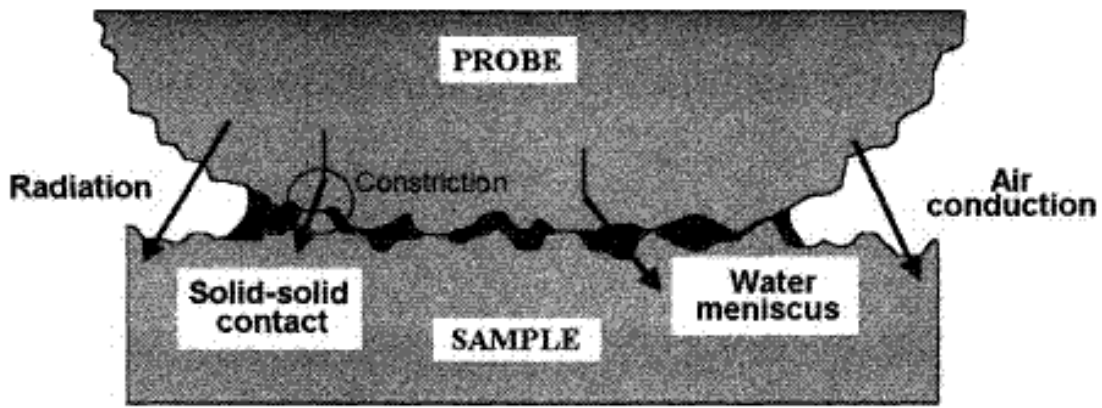

Fig 2:

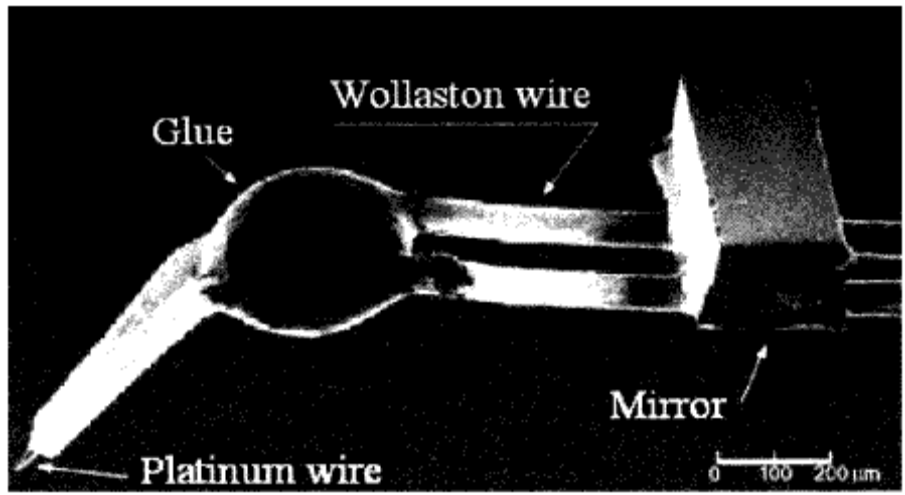

Fig 3:

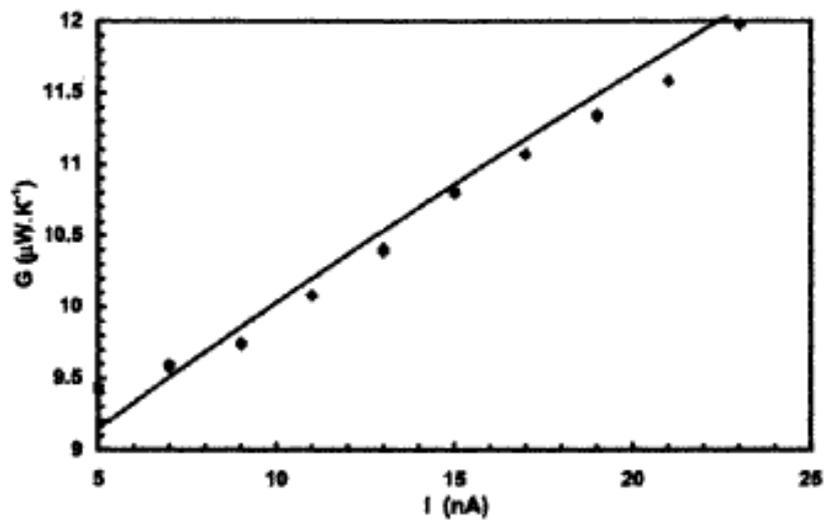

Fig 4:

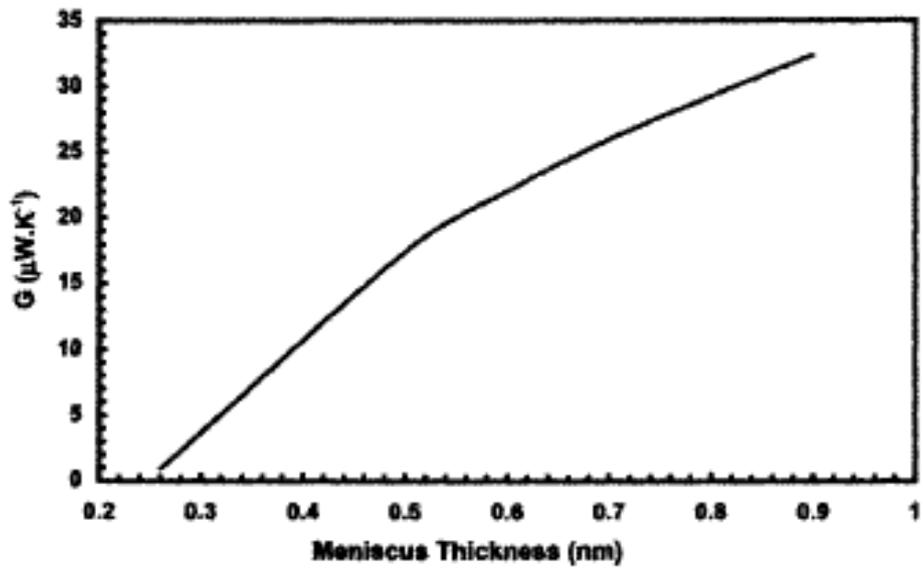

Fig 5:

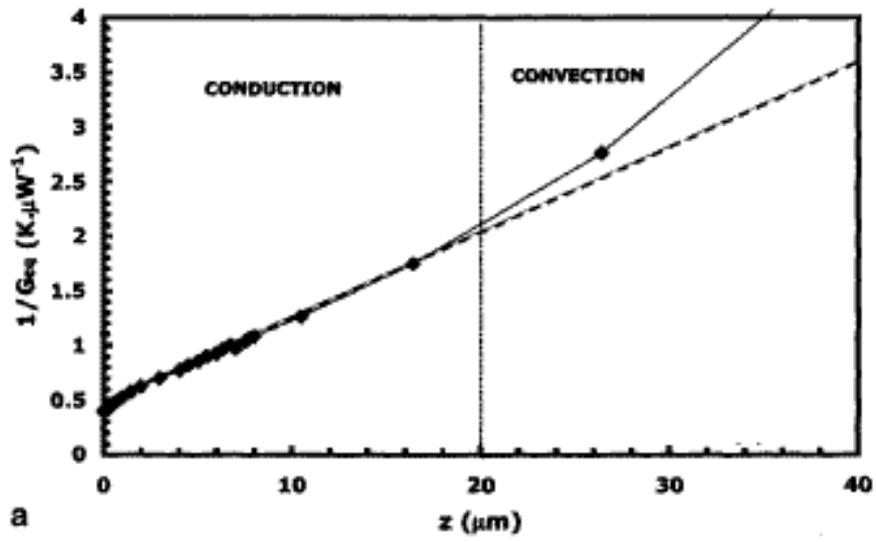

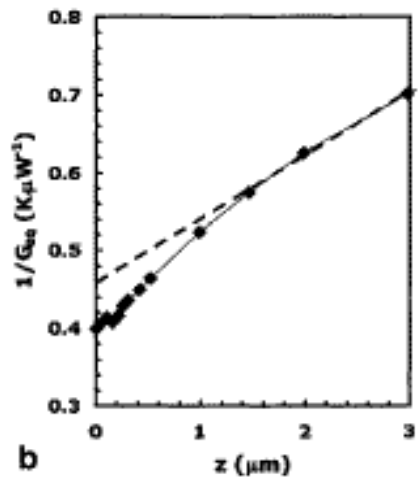

Fig 6:

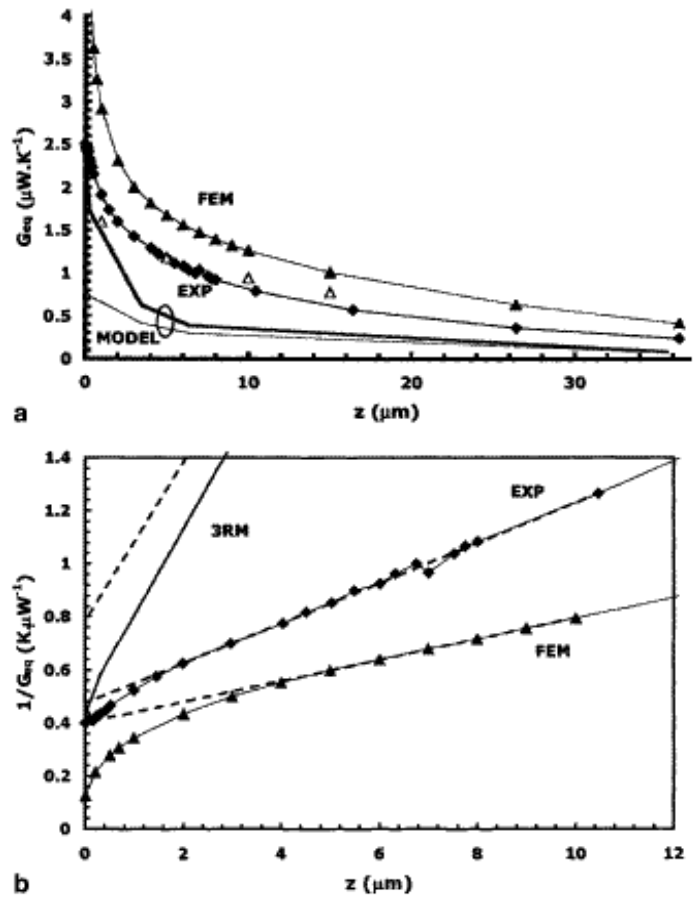

Fig 7:

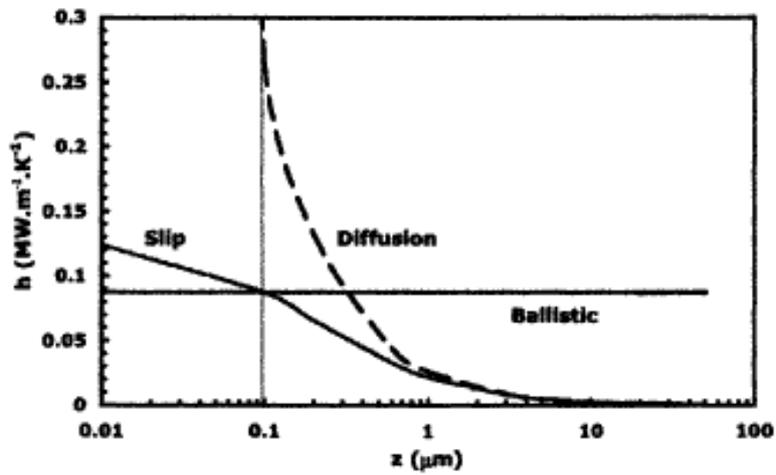

Fig 8:

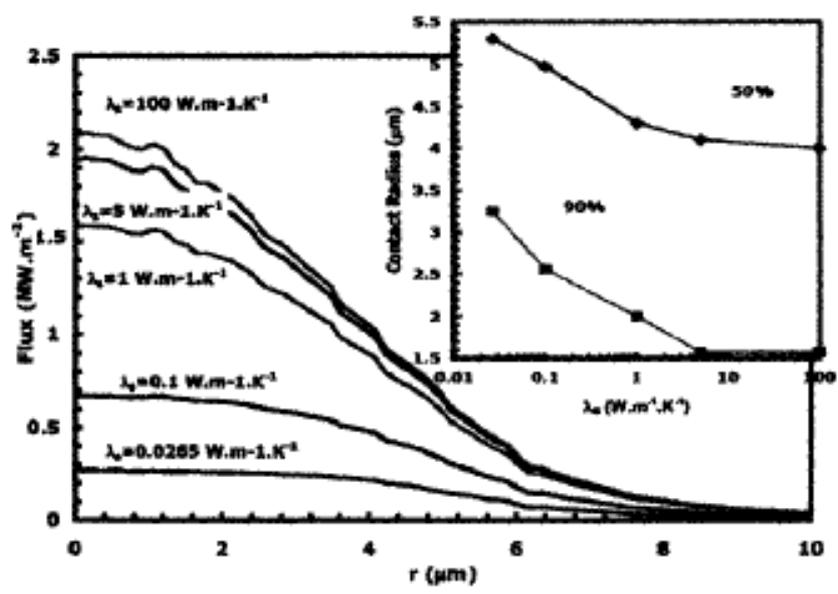